\documentstyle[amstex,amsfonts,aps,prb,twocolumn]{revtex}

\begin{document}
\title{The Rashba Hamiltonian and electron transport}
\author{Laurens W.\ Molenkamp and Georg Schmidt}
\address{Physikalisches Institut (EP3), Universit\"at\\
W\"urzburg, D-97074 W\"urzburg, Germany}
\author{Gerrit E.W.\ Bauer}
\address{Theoretical Physics Group, Department of Applied Physics and DIMES, \\
Delft University of Technology, Lorentzweg 1, 2628 CJ Delft, The Netherlands }
\date{\today }
\maketitle

\begin{abstract}
The Rashba Hamiltonian describes the splitting of the conduction
band as a result of spin-orbit coupling in the presence of an
external field and is commonly used to model the electronic
structure of confined narrow-gap semiconductors. Due to the mixing
of spin states some care has to be exercised in the calculation of
transport properties. We derive the velocity operator for the
Rashba-split conduction band and demonstrate that the transmission
of an interface between a ferromagnet and a Rashba-split
semiconductor does not depend on the magnetization direction, in
contrast with previous assertions in the literature.
\end{abstract}

\pacs{Valid PACS appear here}

Narrow-gap semiconductors, most notably InAs, play an important role in the
rapidly evolving field of spintronics. As non-magnetic element in hybrid
devices these materials are expected to help control the electron spin
states, just like the electron charge is controlled in conventional
electronic devices. Part of this potential stems from the natural
two-dimensional electron gas (2DEG) on clean InAs surfaces, which allows
high-quality ohmic contacts to superconductors and ferromagnets. Another
reason is the seminal paper of Datta and Das,\cite{Datta} which describes
how the electrical field of an external gate electrode can be used to
manipulate the precession of a conduction electron spin. Essential for this
mechanism is the field-dependent spin-orbit coupling, which is relatively
large and well-established for the 2DEG on InAs. It is now generally
accepted that the spin-orbit interaction in narrow-gap 2DEGs is governed by
the Rashba Hamiltonian,\cite{rashba} which increases linearly with the
electron wave vector.

The spin-orbit-interaction induced `spin-splitting' is sometimes confused
with an exchange or Zeeman splitting. However, the latter require breaking
of the time inversion symmetry and are therefore fundamentally different
from the former. It is then not surprising that physical properties like
exciton spin splittings or, in the present context, spin-dependent transport
properties of narrow-gap hybrid devices are not well understood. In a recent
paper, for example, it was argued\cite{dirk} that the conductance of the
interface between a ferromagnet and a spin-orbit spin-split semiconductor
should change on a flip of the magnetization direction of the ferromagnet.
This obviously cannot be correct because in the absence of an external
magnetic field, the spin-quantization axis in the (isotropic) semiconductor
can be rotated with the magnetization direction, which should therefore be
without physical consequences.\cite{stray} The problem with the calculations
of Ref. \onlinecite{dirk} can be traced to an incorrect treatment of the
velocity operator, which in the presence of spin-orbit interaction is not
simply given by $\hbar \vec{k}/m$, where $m$ is the effective mass of an
electron and $\vec{k}$ its wavevector.

It is the purpose of our communication to clarify the issues mentioned
above. First, we will discuss the nature of the eigenstates of the Rashba
Hamiltonian in some detail and derive the proper velocity operator. For
comparison, we give similar expressions for the eigenstates and velocity
operator for a Stoner-Wohlfarth ferromagnet. Finally, we calculate
explicitly the transmission coefficient between a ferromagnet and a
``Rashba-split'' electron gas and show that the contact conductance is
invariant with respect to a magnetization reversal of the ferromagnet.%

The Hamiltonian of an otherwise free electron system, but including the
Rashba spin-orbit scattering term reads:\cite{rashba}
\begin{equation}
H=-\frac{\hbar ^{2}\vec{\nabla}^{2}}{2m}+\alpha \left( -i\vec{\nabla}\times
\vec{E}\right) \cdot {\bf \vec{\sigma}}  \label{RashbaH}
\end{equation}
where $\alpha $ is an effective mass parameter and ${\bf \vec{\sigma}=}%
\left( \sigma _{x,}\sigma _{y},\sigma _{z}\right) $ is the vector of Pauli
spin matrices. For a 2DEG with a confining electric field normal to the
interface $\vec{E}=\left( 0,0,E_{z}\right) :$%
\begin{gather}
H=  \nonumber \\
\left(
\begin{array}{cc}
E_{0}-\frac{\hbar ^{2}}{2m}\left( \frac{\partial ^{2}}{\partial x^{2}}+\frac{%
\partial ^{2}}{\partial y^{2}}\right)  & \left\langle \alpha
E_{z}\right\rangle \left( \frac{\partial }{\partial x}-i\frac{\partial }{%
\partial y}\right)  \\
-\left\langle \alpha E_{z}\right\rangle \left( \frac{\partial }{\partial x}+i%
\frac{\partial }{\partial y}\right)  & E_{0}-\frac{\hbar ^{2}}{2m}\left(
\frac{\partial ^{2}}{\partial x^{2}}+\frac{\partial ^{2}}{\partial y^{2}}%
\right)
\end{array}
\right)
\end{gather}
where $\left\langle \alpha E_{z}\right\rangle $ is the expectation value
over the lowest subband with energy $E_{0}.$ Experimentally, one typically
observes\cite{nitta,stradling} values for $\langle \alpha E_{z}\rangle $ on
the order of $10^{-11}$ eV m. The eigenstates for in-plane motion
(identified by their quantum numbers $\vec{k}=\left( k_{x},k_{y}\right) $
and $s=\pm 1$) are
\begin{equation}
\phi _{\vec{k}s}\left( \vec{r}\right) =N_{\vec{k}s}e^{i\vec{k}\vec{r}}%
{is\left( \frac{k_{x}}{k}-i\frac{k_{y}}{k}\right)  \choose 1}%
.
\end{equation}
where $N_{\vec{k}s}$ is a normalization factor. From this expression for the
eigenstates, it is immediately obvious that the Rashba-split subbands are
not spin-polarized.\cite{limits} The electron energy dispersion relation $E_{%
\vec{k}s}$ reads (see Fig. 1a):
\begin{equation}
E_{\vec{k}s}=E_{0}+\frac{\hbar ^{2}}{2m}\left[ \left( k+sk_{R}\right)
^{2}-k_{R}^{2}\right]
\end{equation}
where $k=\sqrt{k_{x}^{2}+k_{y}^{2}},\;k_{R}=\left\langle \alpha
E_{z}\right\rangle m/\hbar ^{2}.$

The normalization factor of the eigenfunctions can be determined in
different ways. Normalization of the probability distribution $\int d\vec{r}%
\left| \phi _{\vec{k}s}\left( \vec{r}\right) \right| ^{2}=1$ gives $N_{\vec{k%
}s}=1/\sqrt{2S}$ where $S$ is the area of the 2DEG. However, for a
calculation of the transport properties it is more convenient to normalize
the states such that its currents are unity in the transport, say $x,$
direction. To this end we have to compute the expectation value of the
current or velocity operator which, in the presence of the Rashba term, are
not simply proportional to the gradient operator anymore. The proper matrix
representation in spinor space can be derived via the Hamilton equation of
motion:
\begin{equation}
\dot{q}=\frac{\partial H}{\partial p};\;\dot{p}=-\frac{\partial H}{\partial q%
};\;\dot{x}=v_{x}=\frac{\partial H}{\partial p_{x}};\;\dot{p}_{x}=-\frac{%
\partial H}{\partial x}
\end{equation}
The velocity operator in the $x$-direction therefore reads:
\begin{equation}
v_{x}=\frac{1}{\hbar }\left(
\begin{array}{cc}
-i\frac{\hbar ^{2}}{m}\frac{\partial }{\partial x} & i\left\langle \alpha
E_{z}\right\rangle \\
-i\left\langle \alpha E_{z}\right\rangle & -i\frac{\hbar ^{2}}{m}\frac{%
\partial }{\partial x}
\end{array}
\right) .
\end{equation}
Requiring $\left\langle \phi _{\vec{k}s}\left( \vec{r}\right) |\hat{v}%
_{x}|\phi _{\vec{k}s}\left( \vec{r}\right) \right\rangle =1$ we find
\begin{equation}
N_{\vec{k}s}=\sqrt{\frac{m}{2\hbar }}\sqrt{\frac{1}{\left| k_{x}\left( 1+s%
\frac{k_{R}}{k}\right) \right| }}
\end{equation}
This value diverges when the group velocity vanishes, {\it i.e.} for $s=-1$
at $k=k_{R}.$

The above considerations for a 2DEG are only slightly modified for a quantum
wire. For the lowest mode:
\begin{align}
H\left( x,p_{x}\right) & =\left(
\begin{array}{cc}
E_{0}^{\prime }+\frac{p_{x}^{2}}{2m} & i\frac{\left\langle \alpha
E_{z}\right\rangle }{\hbar }p_{x} \\
-i\frac{\left\langle \alpha E_{z}\right\rangle }{\hbar }p_{x} &
E_{0}^{\prime }+\frac{p_{x}^{2}}{2m}
\end{array}
\right) \\
v_{x}& =\frac{\partial H}{\partial p_{x}}=\left(
\begin{array}{cc}
\frac{p_{x}}{m} & i\frac{\left\langle \alpha E_{z}\right\rangle }{\hbar } \\
-i\frac{\left\langle \alpha E_{z}\right\rangle }{\hbar } & \frac{p_{x}}{m}
\end{array}
\right)
\end{align}
Note that there is no Rashba level splitting in a quantum dot.

It is instructive to compare the Rashba Hamiltonian with that of a 2D
non-collinear ferromagnet with a dispersion as sketched in Fig. 1b:
\begin{equation}
H_{ncf}=\left(
\begin{array}{cc}
\frac{p_{x}^{2}}{2m}+\frac{p_{y}^{2}}{2m} & 0 \\
0 & \frac{p_{x}^{2}}{2m}+\frac{p_{y}^{2}}{2m}
\end{array}
\right) +\Delta U^{+}\sigma _{z}U
\end{equation}
where $2\Delta $ is the exchange splitting and
\begin{equation}
U\left( \theta ,\varphi \right) =\left(
\begin{array}{cc}
\cos \theta /2 & e^{-i\varphi }\sin \theta /2 \\
e^{i\varphi }\sin \theta /2 & -\cos \theta /2
\end{array}
\right)
\end{equation}
is a unitary rotation matrix corresponding to a magnetization direction of $%
\vec{m}=\left( \sin \theta \cos \varphi ,\sin \theta \sin \varphi ,\cos
\theta \right) .$ For plane wave states with wave vector $\vec{k}$ and $\vec{%
m}=\left( 0,-1,0\right) $ the Hamiltonian for the ferromagnet
\begin{equation}
H_{ncf}\left( k,\theta =\frac{\pi }{2},\varphi =-\frac{\pi }{2}\right)
=\left(
\begin{array}{cc}
\frac{\hbar ^{2}}{2m}k^{2} & i\Delta \\
-i\Delta & \frac{\hbar ^{2}}{2m}k^{2}
\end{array}
\right)
\end{equation}
is formally equivalent to that of the Rashba Hamiltonian
\begin{equation}
H_{R}=\left(
\begin{array}{cc}
\frac{\hbar ^{2}}{2m}k^{2} & i\Delta _{R} \\
-i\Delta _{R}^{\ast } & \frac{\hbar ^{2}}{2m}k^{2}
\end{array}
\right)
\end{equation}
with $\Delta _{R}=\left\langle \alpha E_{z}\right\rangle \left(
k_{x}-ik_{y}\right) .$ In the ferromagnet the velocity operator is always
diagonal in spin space, however:
\begin{equation}
v_{x,ncf}\left( k,\theta ,\varphi \right) =\frac{p_{x}}{m}\left(
\begin{array}{cc}
1 & 0 \\
0 & 1
\end{array}
\right) .
\end{equation}

In order to demonstrate explicitly that transport through a Rashba
semiconductor/ferromagnet junction does not depend on the magnetization
direction of the ferromagnet, it is sufficient to consider the simple case
of a single mode quantum point contact (Fig. 2) without an additional
interface potential barrier. A ferromagnet on the left side of the contact
(its electronic states will be indicated by superscript ${\frak L}$ in the
following) is attached (at $x=0$) to a Rashba semiconductor on the right
(superscript ${\frak R}$). In the semiconductor we have eigenstates at the
Fermi energy $E_{F}=\frac{\hbar ^{2}}{2m}k_{F}^{2}$ at wave vectors $k_{s}\
=-sk_{R}+\sqrt{k_{R}^{2}+k_{F}^{2}}$ which are taken to be positive in the
following. The states at the Fermi energy are right moving:
\begin{equation}
\phi _{k_{+}}^{{\frak R}}\left( x\right) =Ne^{ik_{+}x}%
{1 \choose -i}%
;\;\phi _{k_{-}}^{{\frak R}}\left( x\right) =Ne^{ik_{-}x}%
{1 \choose i}%
\end{equation}
and left moving
\begin{equation}
\phi _{-k_{+}}^{{\frak R}}\left( x\right) =Ne^{-ik_{+}x}%
{1 \choose i}%
;\;\phi _{-k_{-}}^{{\frak R}}\left( x\right) =Ne^{-ik_{-}x}%
{1 \choose -i}%
\end{equation}
with normalization
\begin{equation}
N=\sqrt{\frac{m}{2\hbar }}\sqrt{\frac{1}{k_{s}+sk_{R}}}=\sqrt{\frac{m}{%
2\hbar }}\sqrt{\frac{1}{\sqrt{k_{R}^{2}+k_{F}^{2}}}}
\end{equation}
The flux normalization reflects the identical group velocities for the two
bands. The normalization is invariant under a unitary transformation which
diagonalizes the Hamiltonian. We have seen above that we can interpret the
Rashba semiconductor as a pseudo-ferromagnet in which the magnetization is
rotated from the $z$ to the $-y$ direction, and with a $k$-dependent
exchange splitting $\Delta _{R}.$ We simplify the situation by taking the
quantization axis of the ferromagnet parallel to the pseudo-magnetization of
the Rashba Hamiltonian by transforming the Rashba Hamiltonian as follows:
\begin{gather}
U^{+}\left( \frac{\pi }{2},-\frac{\pi }{2}\right) H_{R}U\left( \frac{\pi }{2}%
,-\frac{\pi }{2}\right) =  \nonumber \\
\left(
\begin{array}{cc}
\frac{p_{x}^{2}}{2m} & 0 \\
0 & \frac{p_{x}^{2}}{2m}
\end{array}
\right) +\left\langle \alpha E_{z}\right\rangle \frac{\partial }{\partial x}%
\left(
\begin{array}{cc}
0 & 1 \\
-1 & 0
\end{array}
\right) ,
\end{gather}
yielding the following eigenstates along the quantization ($-y$) axis of the
ferromagnet
\begin{align}
U\phi _{k_{+}}^{{\frak R}}\left( x\right) & =\sqrt{\frac{m}{\hbar \sqrt{%
k_{R}^{2}+k_{F}^{2}}}}%
{1 \choose 0}%
, \\
U\phi _{k-}^{{\frak R}}\left( x\right) & =\sqrt{\frac{m}{\hbar \sqrt{%
k_{R}^{2}+k_{F}^{2}}}}%
{0 \choose 1}%
.
\end{align}
On the left side we assume first a half-metallic ferromagnet{\it , }for
which the conduction electrons are either all spin up or down with wave
vector $k_{F}$:
\begin{eqnarray}
\phi _{_{\uparrow }}^{{\frak L}}\left( x\right) &=&\sqrt{\frac{m}{\hbar k_{F}%
}}e^{ik_{F}x}%
{1 \choose 0}%
\\
\phi _{\downarrow }^{{\frak L}}\left( x\right) &=&\sqrt{\frac{m}{\hbar k_{F}}%
}e^{ik_{F}x}%
{0 \choose 1}%
\end{eqnarray}
Assuming that the spin is up on the left side, we can now write the
eigenstates of the ferromagnet in terms of the reflection coefficient $%
r_{\uparrow }$:
\begin{equation}
\chi _{\uparrow }^{{\frak L}}\left( x\right) =\sqrt{\frac{m}{\hbar k_{F}}}%
\left[ e^{ik_{F}x}%
{1 \choose 0}%
+r_{\uparrow }e^{-ik_{F}x}%
{1 \choose 0}%
\right] .
\end{equation}
On the side of the Rashba-split semiconductor, we have transmission for one
spin direction only, which corresponds to a wave vector $k_{+}.$
\begin{equation}
\chi _{\uparrow }^{{\frak R}}\left( x\right) =t_{\uparrow +}\sqrt{\frac{m}{%
\hbar \sqrt{k_{R}^{2}+k_{F}^{2}}}}e^{ik_{+}x}%
{1 \choose 0}%
.
\end{equation}
where $t_{\uparrow +}$ is the transmission coefficient. The transport
coefficients are determined by the requirement of the continuity of the wave
function and its flux ({\em not simply the derivative}) at the interface $%
x=0 $:
\begin{align}
\chi _{\uparrow }^{{\frak L}}\left( 0\right) & =\chi _{\uparrow }^{{\frak R}%
}\left( 0\right) \\
v_{x}\chi _{\uparrow }^{{\frak L}}\left( x\right) |_{x=0}& =v_{x}\chi
_{\uparrow }^{{\frak R}}\left( x\right) |_{x=0}
\end{align}
The condition for flux continuity can be rewritten as
\begin{gather}
\ \frac{\hbar }{im}\frac{\partial }{\partial x}\chi _{\uparrow }^{{\frak L}%
}\left( x\right) |_{x=0}  \nonumber \\
=\frac{\hbar }{im}\left( \frac{\partial }{\partial x}+ik_{R}\right)
t_{\uparrow }\sqrt{\frac{m}{\hbar \sqrt{k_{R}^{2}+k_{F}^{2}}}}%
e^{ik_{+}x}|_{x=0} \\
=\frac{\hbar }{m}\left( k_{+}+k_{R}\right) =\frac{\hbar }{m}t_{\uparrow }%
\sqrt{k_{R}^{2}+k_{F}^{2}}  \label{t1}
\end{gather}
We can now calculate the conductance via the Landauer formula:
\begin{equation}
G_{\uparrow }=\frac{e^{2}}{h}\left| t_{\uparrow }\right| ^{2}=\frac{e^{2}}{h}%
\frac{4\sqrt{1+\left( k_{R}/k_{F}\right) ^{2}}}{\left( 1+\sqrt{1+\left(
k_{R}/k_{F}\right) ^{2}}\right) ^{2}}
\end{equation}
To calculate $G_{\downarrow }$, we flip the magnetization of the ferromagnet
on the left side, yielding as incoming state
\begin{equation}
\chi _{\downarrow }^{{\frak L}}\left( x\right) =\sqrt{\frac{m}{\hbar k_{F}}}%
\left[ e^{ik_{F}x}%
{0 \choose 1}%
+r_{\downarrow }e^{-ik_{F}x}%
{0 \choose 1}%
\right]
\end{equation}
while transmission occurs only into
\begin{equation}
\chi _{\downarrow }^{{\frak R}}\left( x\right) =t_{\downarrow }\sqrt{\frac{m%
}{\hbar \sqrt{k_{R}^{2}+k_{F}^{2}}}}e^{ik_{-}x}%
{0 \choose 1}%
\end{equation}
Flux continuity gives:
\begin{align}
v_{x}\chi _{\downarrow }^{{\frak L}}\left( x\right) |_{x=0}& =\frac{\hbar }{%
im}\frac{\partial }{\partial x}\chi _{\downarrow }^{{\frak L}}\left(
x\right) |_{x=0}=v_{x}\chi _{\downarrow }^{{\frak R}}\left( x\right) |_{x=0}
\\
& =t_{\downarrow }\frac{\hbar }{m}\left( k_{-}-k_{R}\right) =\frac{\hbar }{m}%
t_{\downarrow }\sqrt{k_{R}^{2}+k_{F}^{2}}
\end{align}
and, comparing this expression with Eq. (\ref{t1}), we see that the
transmission coefficient is identical for up and down spins. This is in
contrast with the counterintuitive results of Ref. \onlinecite{dirk} -
where, we believe, an incorrect velocity operator has been applied.

Since the effective-mass Rashba Hamiltonian of Eq. (1) is isotropic, the
interface conductance is invariant under arbitrary rotations of the
magnetization direction. In addition, the above calculations may be
generalized to transmission from a weak ferromagnet with\ both spins
occupied up to Fermi numbers $k_{\uparrow /\downarrow }$ with
\begin{equation}
G=\frac{e^{2}}{h}\sum_{\sigma =\uparrow ,\downarrow }\frac{4\sqrt{%
k_{R}^{2}+k_{F}^{2}}}{\left( 1+\sqrt{\left( k_{R}^{2}+k_{F}^{2}\right)
/k_{\sigma }^{2}}\right) ^{2}}
\end{equation}
The interface conductance should therefore not affect anisotropies due to
interference effects in the Datta transistor.\cite{Datta}%

We hope that this paper will help to dispel the confusion concerning the
transport properties of semiconductors with spin-orbit interactions. We
compared eigenstates and velocity operators for two systems, a non-magnetic
2DEG in the presence of the Rashba Hamiltonian and a non-collinear
Stoner-Wohlfarth model ferromagnet. As expected, the transmission
coefficient of an interface between a ferromagnet and a Rashba-split
semiconductor is found independent on the magnetization direction of the
ferromagnet.

Note that the independence of the total conductance of a single
ferromagnetic/normal metal interface on the magnetization direction is quite
general, but does not mean that the interface is not spin-selective. Indeed,
a ferromagnet does inject a net spin into the nonmagnetic material, with
efficiencies that depend on the specific electronic band structures.\cite
{ferr,kirc} Small modulations of a single interface conductance could be
achieved in principle by forcing the magnetization vector of the ferromagnet
into directions which deviate from the crystal symmetry axes. However, in
order to detect a strongly spin-polarized interface transmission by a
transport experiment, an analyzing ferromagnet is essential. This is
employed, of course, in the giant magnetoresistance effect. In
semiconductors, the spin-polarized current can also be detected by the
circular optical polarization of the electroluminescence of a light emitting
diode.\cite{spinled}

We thank the Volkswagenstiftung for supporting a R\"{o}ntgen-Professorship
of G.E.W.B. at W\"{u}rzburg University. G.E.W.B. acknowledges support by the
NEDO project NTDP-98; L.W.M. and G.S. acknowledge support by the DARPA SPINS
programme.

{\it Note added:} After submission of this manuscript, we received
a preprint by Z\"{u}licke and Schroll with similar results. Bruno
and Pareek, cond-mat/0105506, report numerical calculations for
the same system. In contrast to what we report here, the latter
authors find a small anisotropy in the transport as a function of
the magnetization angle. These anisotropies are allowed by
(Casimir-Onsager) symmetry, but they vanish for the effective mass
Hamiltonian (\ref{RashbaH}), which does not contain any
``warping'' corrections which reflect the reduced symmetry of the
crystalline lattice.

\begin{figure}[h]
\caption{Schematic representation of the conduction band structure of (a) a
semiconductor in which the spin-degeneracy is broken by spin-orbt
interaction as described by the Rashba Hamiltonian and (b) a spin-polarized
exchange-split band ferromagnet (Stoner-Wohlfarth model). }
\label{fig1}
\end{figure}
%

%
\begin{figure}[h]
\caption{The system under consideration: we discuss the conductance of a
ferromagnet-semiconductor hybrid quantum point contact. The band structures
of the two materials are depicted in Fig. 1. The constriction separating
the two materials symbolizes the single-channel adiabatic transport we
assume in our calculation. The
ferromagnet is magnetized in the $-y$-direction and the current flows
in the $x-$direction.}
\label{fig2}
\end{figure}

%


\begin{references}
\bibitem{stray}  This argument holds also in the presence of stray fields
from the ferromagnet which rotate invariantly with the magnetization.
\bibitem{limits}  Note that that even in the limit of $\left\langle \alpha
E_{z}\right\rangle =0$ the eigenstates (3) are linear combinations of
spin-up and spin-down states. Pure spin eigenstates could be obtained in
this limit by including a Zeeman spin splitting $\Delta $ in the Hamiltonian
which in the end is allowed to vanish.
\end{references}
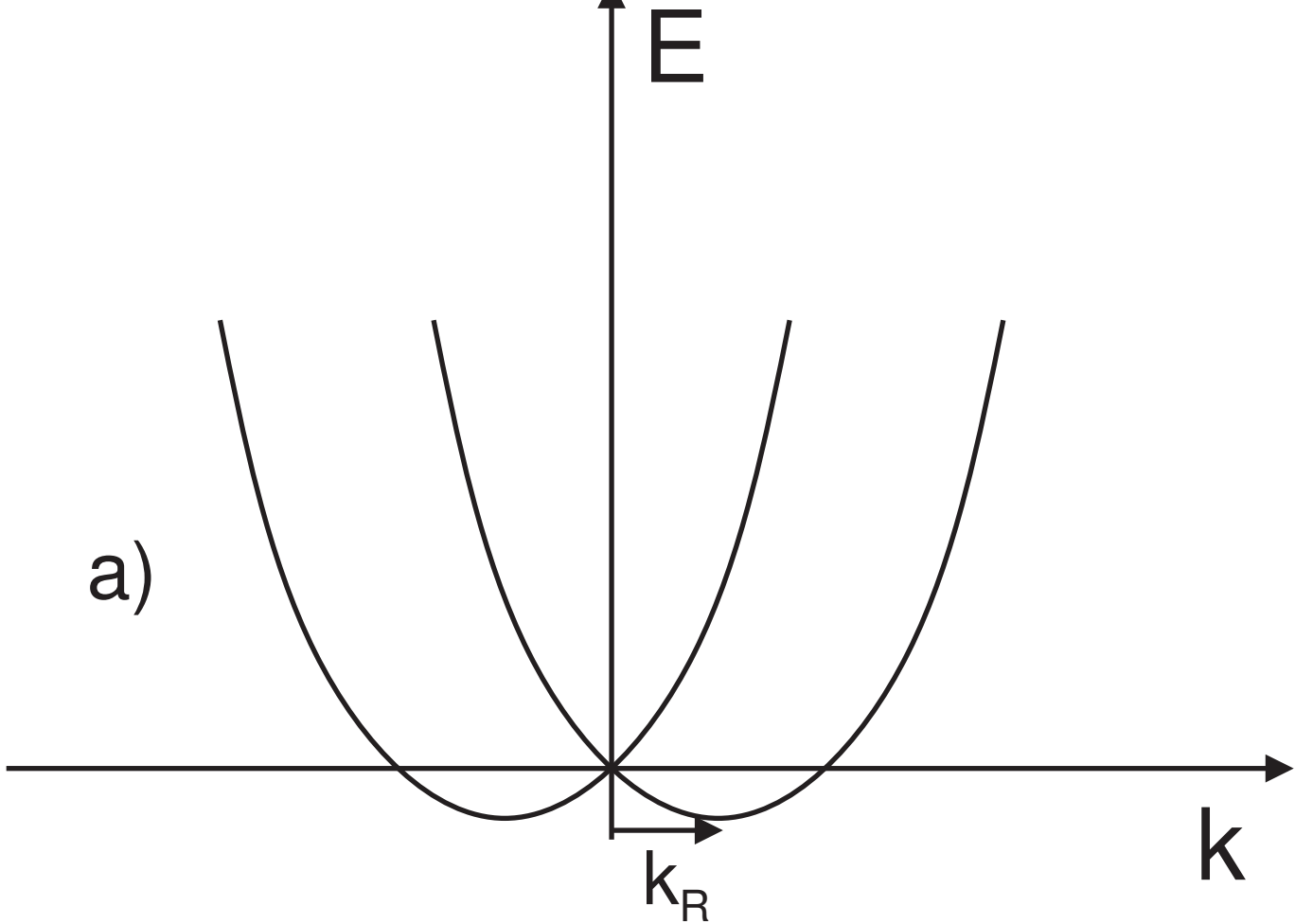
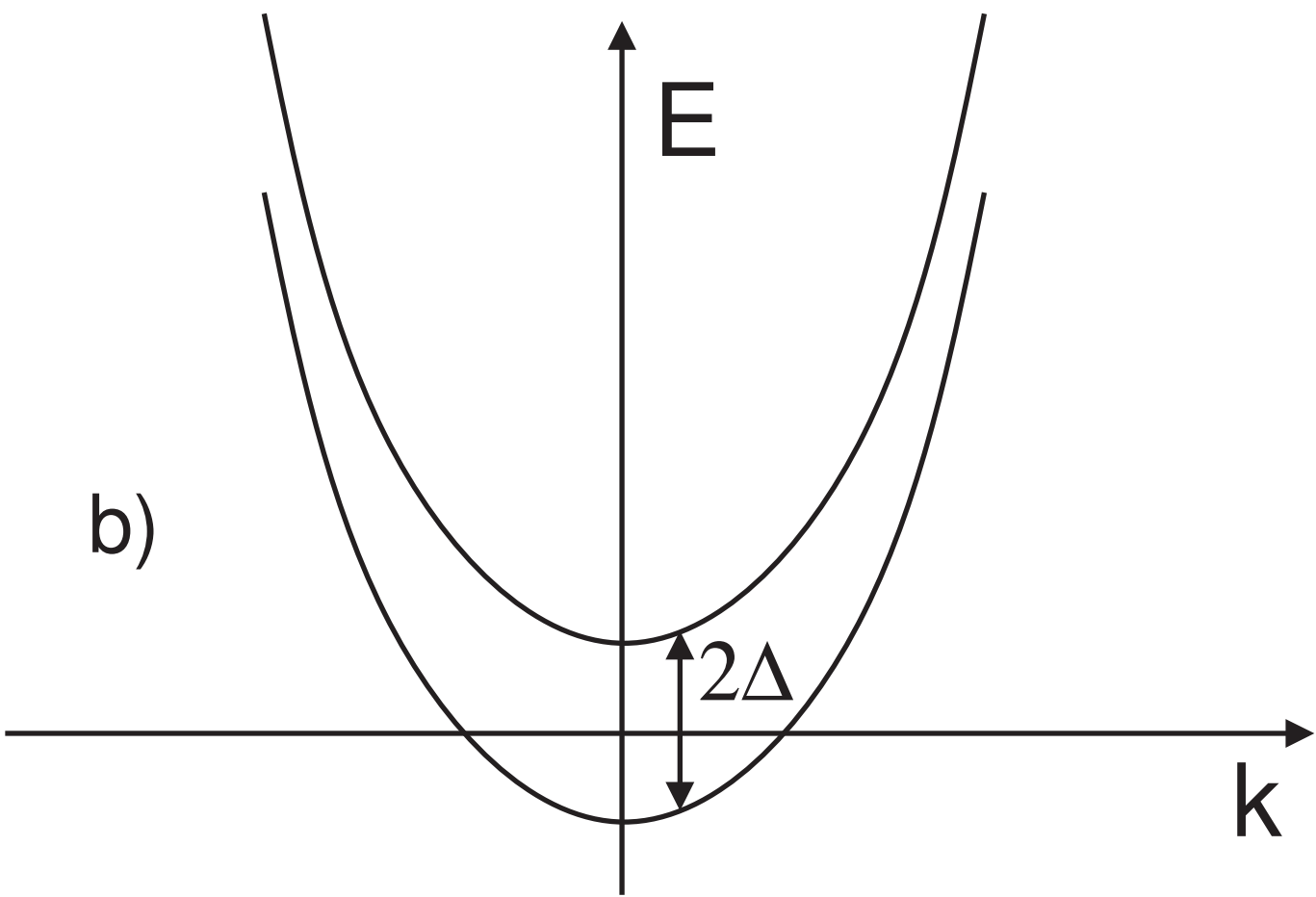
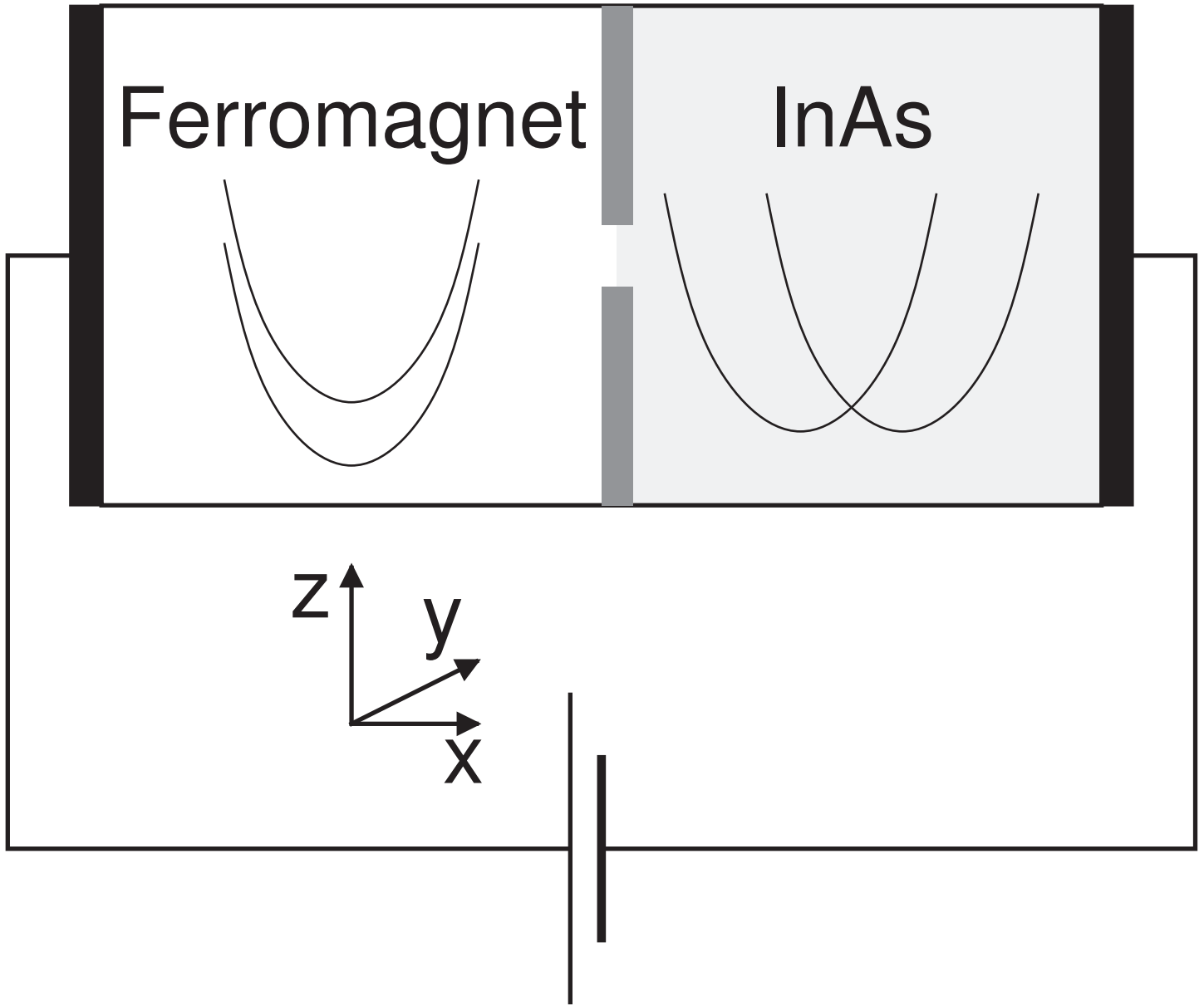
\end{document}